\documentclass{optica-article}

\journal{opticajournal} 

\articletype{Research Article}

\usepackage{lineno}

\usepackage{hyperref}
\hypersetup
{
colorlinks,%
citecolor=blue,%
linkcolor=black,%
urlcolor=black,%
}

  \newcommand{\I}{\mathrm{i}}

  \newcommand{\smlket}[1]{| #1 \rangle}
  
  \newcommand{\smlbraket}[2]{\langle #1 | #2 \rangle}
  \newcommand{\smlketbra}[2]{|#1 \rangle \langle#2 |}

  \newcommand{\ket}[1]{\left| #1 \right\rangle}
  \newcommand{\bra}[1]{\left\langle #1 \right|}
  
  \newcommand{\ketbra}[2]{\left |#1 \middle\rangle \middle\langle#2 \right|}

  \newcommand{\nt}{\negthickspace}
  \newcommand{\sket}[1]{| #1  \rangle \nt \rangle}
  \newcommand{\sbra}[1]{\langle \nt \langle #1 |}

  \newcommand{\abs}[1]{\left| #1 \right|}
  \newcommand{\argp}[1]{\left( #1 \right)}
  \newcommand{\args}[1]{\left[ #1 \right]}
  \newcommand{\argc}[1]{\left\{ #1 \right\}}

\begin{document}

\title{Error suppression in multicomponent cat codes with photon subtraction and teleamplification}

\author{Saurabh U. Shringarpure\authormark{1}, Yong Siah Teo\authormark{2}, and Hyunseok Jeong\authormark{3}}

\address{Department of Physics $\&$ Astronomy, Seoul National University, 1 Gwanak-ro, Gwanak-gu, Seoul 08826, Republic of Korea}

\email{\authormark{1}saurabh.s@snu.ac.kr, \authormark{2}ys\_teo@snu.ac.kr, \authormark{3}h.jeong37@gmail.com} 


\begin{abstract*} 
It is known that multiphoton states can be protected from decoherence due to a passive loss channel by applying noiseless attenuation before and noiseless amplification after the channel. In this work, we propose the combined use of multiphoton subtraction on four-component cat codes and teleamplification to effectively suppress errors under detection and environmental losses. The back-action from multiphoton subtraction modifies the encoded qubit encoded on cat states by suppressing the higher photon numbers, while simultaneously ensuring that the original qubit can be recovered effectively through teleamplification followed by error correction, thus preserving its quantum information. With realistic photon subtraction and teleamplification-based scheme followed by optimal error-correcting maps, one can achieve a worst-case fidelity (over all encoded pure states) of over $93.5\%$ ($82\%$ with only noisy teleamplification) at a minimum success probability of about $3.42\%$, under a $10\%$ environmental-loss rate, $95\%$ detector efficiency and sufficiently large cat states with the coherent-state amplitudes of 2. This sets a promising standard for combating large passive losses in quantum-information tasks in the noisy intermediate-scale quantum (NISQ) era, such as direct quantum communication or the storage of encoded qubits on the photonic platform.

\end{abstract*}

\section{Introduction}
Multiphoton superposition states, crucial for various quantum applications, are susceptible to decoherence because of photon loss, particularly affecting higher energy terms. Noiseless attenuation and \mbox{noiseless amplification} have been shown to be effective in suppressing decoherence due to photon loss~\cite{mivcuda2012noiseless}. These nondeterministic techniques work by strategically reducing the probability amplitudes of higher photon number terms suffering from loss. 

Bosonic error-correcting codes offer an alternative for protecting qubits from the impact of photon loss. Quantum error correction using these codes has recently garnered attention due to their physical-resource efficiency, as opposed to discrete-variable counterparts that require coherence over a large number of multiple \mbox{physical modes~\cite{albert2018performance,terhal2020towards,joshi2021quantum,cai2021bosonic}.} Numerically optimized bosonic codes tailored for loss channels have been explored in previous studies \cite{michael2016new, albert2018performance} where entanglement fidelity of the channels is used as a figure of merit. Reference \cite{leviant2022quantum} extends this analysis to include general loss-dephasing channels, where a key takeaway was that for large values of noise, encodings with higher average energy may not be suitable. 

In this context, we explore the interplay between reducing average energy and ensuring that an encoded qubit can be recovered by error correction. We also consider the influence of additional noise introduced during realistic error suppression on the encoded states. 
Optimized encodings for a given channel mentioned earlier may not be easy to realize in practice. 

For example, Gottesman–Kitaev–Preskill (GKP) \cite{gottesman2001encoding} codes with high squeezing, robust against loss channels pose significant experimental challenges. Nevertheless, there has recently been some progress in generating them on \mbox{photonic platforms}~\cite{dahan2023creation, konno2023propagating}. This also motivates error suppression with currently feasible encodings. This paper focuses on multicomponent cat codes, particularly four-component cat codes that are resilient to errors induced by single-photon annihilation \cite{mirrahimi2014dynamically, ofek2016extending, hastrup2020deterministic, su2022universal, hastrup2022all} and restricts to the amplitude-damping or loss channel, a predominant noise source across \mbox{various platforms.} We study this on photonic platforms, where \mbox{noiseless attenuation} is generally implemented using zero-photon subtraction (ZPS) \cite{nunn2021heralding,nunn2022modifying,nunn2023transforming} and has been shown to preserve the coherence of two-component \mbox{cat states \cite{shringarpure2022coherence}.} 

Subtraction of an arbitrary number of photons from a cat state results in another cat state, while also implementing noiseless attenuation like ZPS. Subtraction of a larger number of photons also gives a higher success probability in many situations. We discuss in some detail the consequences of this for fidelity and success rates. Alternatives to error suppression of the current discussion, where deterministic squeezing is considered instead of multiphoton subtraction, have been \mbox{explored~\cite{le2018slowing,brewster2018reduced,park2022slowing, schlegel2022quantum}.} 
Recent demonstrations, such as those in Ref. \cite{pan2023protecting} on the \mbox{circuit-QED} platform, encourage interest in active preemptive error suppression. 
Error filtration \cite{gisin2005error, lee2023error, miguel2023superposed} and channel correction~\cite{slussarenko2022quantum} also seek to protect quantum information by suppressing errors. 

In this work, we similarly analyze a continuous-variable (CV) system that encodes a single qubit and exploit the rotation symmetry of the cat codes. Recently, quantum error mitigation with rotationally symmetric bosonic codes was explored in Ref. \cite{endo2022quantum} which also supports this work. However, we focus on implementing the noiseless loss suppression scheme for photonic platforms and take into account the additional sources of noise that would be involved in \mbox{its realization.} Error mitigation techniques that suppress errors in expectation values during data postprocessing have gained recent attention~\cite{cai2023quantum, li2017efficient, temme2017error}, but these are different from active error suppression methods that herald {\it{event-ready states}} of interest here. In these techniques, the quantum state is destroyed during measurements, whereas we would like to preserve the quantum information for later~applications. 

Deterministic phase-preserving linear amplification always adds noise \cite{caves1982quantum, caves2012quantum}. Efficient and noiseless amplification of two-component cat codes become feasible, albeit nondeterministically, with multiple copies of the state as a resource \cite{lund2004conditional,jeong2005production, oh2018efficient}. Quantum scissors have been considered for the approximate \mbox{noiseless amplification} of low-amplitude states \cite{ralph2009nondeterministic, barbieri2011nondeterministic}. Subsequently, teleportation-based noiseless linear amplifiers were introduced for multicomponent cat codes that work exactly \cite{neergaard2013quantum,fiuravsek2022teleportation, guanzon2023noiseless}. We adopt this approach to amplify the amplitudes of the cat states specifically before recovery and generalize the standard teleamplification to herald on a larger set of photon number detections along with multiphoton subtraction for the noiseless~attenuation. The recovery for each compatible outcome is appropriately adapted as shown in Fig.~\ref{fig: prelims}(e).

In our analyses, we prefer to use the worst-case fidelity across all input pure states to characterize the performance. We find that a worst-case fidelity of over $93.5\%$ can be achieved, at a minimum success probability of about $3.42\%$, under a $10\%$ environmental-loss rate, a realistic $95\%$ detector efficiency, for cat states with the coherent-state amplitudes of as large as 2. The ability to protect such large cat states is quite appealing for quantum information applications in the near term, wherein large passive losses before error correction are unavoidable. Storing photonic qubits in a cavity or long fiber loops, or direct communication of the qubits over long distances provide some near-term examples. These contribute to the development of a distributed quantum~network~\cite{azuma2023quantum}.

The rest of the paper is organized as follows. The preliminaries of photon subtraction, teleamplification, photon losses, and multicomponent cats with some of the notations used throughout are presented in \mbox{Sec. \ref{sec: prelims}}.
Section \ref{sec: error suppression} discusses the error suppression for four-component cat codes.
Optimal recoveries in Sec.~\ref{sec: optimal-recovery} and the results are presented in Sec. \ref{sec: results}. The paper ends with a discussion \mbox{in Sec.~\ref{sec: discussion}.}

\section{Preliminaries}
\label{sec: prelims}

\begin{figure}
\centering
\includegraphics[width=7cm]{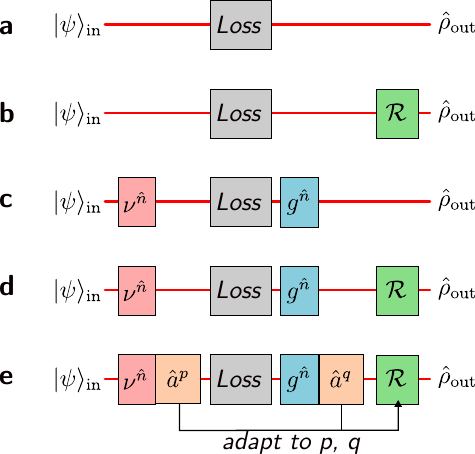}
\caption{(a) A pure photonic state passing through a photon loss channel becomes a mixture. Higher photon number terms in the initial state are more likely to be lost to the environment. (b) The quantum information encoded in the error-correcting codes is recovered after loss. The encoded states are typically a large average photon number. (c) Quantum information is made resilient to noise from photon loss by noiseless loss suppression. Operations with $\nu<1$ and $g>1$ constitute noiseless attenuation and amplification processes, respectively. It works by suppressing large energy terms. (d) The two approaches are combined, where the states encoded in the error-correcting codes are passed through a loss suppression procedure, which also suppresses uncorrectable logical errors on the encoded states. (e) The procedure is extended to include arbitrary photon annihilations during noiseless attenuation and amplification sections. Error correction is suitably adapted to outcomes obtained during the general photon subtraction and teleamplification. } 
\label{fig: prelims}
\end{figure}

Passive photon loss affects the superpositions of the input Fock states $\ket{\psi_{\rm in}}=\sum_{n=0}^{\infty}{c_n\ket{n}}$ by decohering them into mixtures $\hat{\rho}_{\rm out}$, as shown in Fig. \ref{fig: prelims}(a).
CV error-correcting codes are multiphoton states with superpositions containing a large number of photons. 
The purpose of these codes is to encode small logical quantum information, such as a single (or a few) qubit(s), on a large number of subsystems. 
Therefore, the infinitely many Fock states are well suited to this end, and this is the reason behind the physical-resource efficiency of CV \mbox{error correction.} As the photon losses occur, they cause errors on the encoded information. 
However, error-correcting codes allow the detection and correction of such errors through a recovery channel $\mathcal{R}$. This includes measuring the syndrome such as the generalized photon number parity of the affected state without destroying the logical superposition, as shown in Fig. \ref{fig: prelims}(b). 

Different encodings are suited to different kinds of noise, but no realistic encoding can detect and correct for any arbitrary number of photon losses, and therefore we naturally end up in a mixed state in the \mbox{computational basis.}
The challenge of error correction is to detect and correct errors faster than they can grow, and there has been great focus in this direction. 

One way to protect the purity of states is the relatively less known method of noiseless loss suppression, shown in Fig. \ref{fig: prelims}(c). 
In this scheme, noiseless attenuation is applied to the initial pure state which is an operator $\nu^{\hat{n}}$ with the constant factor \mbox{$\nu<1$}. It results in a state $\propto \sum_{n=0}^{\infty}c_n\nu^n\ket{n}$.
Larger $n$ terms are thus exponentially suppressed, and thus the state is greatly protected against the occurrence of photon losses.

Noiseless attenuation is a nonunitary operation and can only be performed nondeterministically. 
Although it makes the states resilient to possible losses, the redistribution of the probability amplitudes associated with the constituent Fock states can drastically distort the input.
The state can be recovered by an effective \textit{inverse} under noiseless amplification which is simply the operation $g^{\hat{n}}$ with the factor $g>1$. If $g=1/\nu$, then $(\nu g)^{\hat{n}}=\hat{I}$. 

In this work, we combine the two approaches to protecting fragile quantum information against decoherence from photon loss, as shown in Fig. \ref{fig: prelims}(d).
We are interested in noiselessly attenuating the physical states of arbitrary computational information encoded in CV encodings before it experiences loss. 
This will naturally distort the state and take it out of the codespace. It is not obvious at the outset if the encoded states are resilient to such nonunitary distortion, photon loss, and after  noiseless amplification, still be able to detect and correct logical errors during recovery.  

In this work, we find that cat encodings due to their rotational symmetric structure are indeed robust under this procedure and can utilize the advantages of error correction and noiseless loss suppression.
Moreover, the particular implementations of noiseless loss suppression for cat states, with photon subtraction and teleamplification, also exhibit another advantage in that an arbitrary number of photon annihilations over these operations, $\nu^{\hat{n}}$ and $g^{\hat{n}}$, do not affect the encoding. This will be explained in detail in Sec. \ref{sec: error suppression}.

These photon annihilations are typically an undesirable consequence of realizing noiseless loss suppression and lead to many failed events, for example, when \mbox{noiseless attenuation} is heralded only on the detection of zero photons. 
However, due to this feature of the cat encoding, we can adapt the recovery channel $\mathcal{R}$ over a larger set of events than would be possible if we restricted ourselves to ZPS and the standard teleamplificaiton, as illustrated by Fig. \ref{fig: prelims}(e).

In the following, preliminaries of photon subtraction, teleamplification, photon loss channel, and multicomponent cat states are presented. This section also introduces the notation we use throughout the paper.

\subsection{Multiphoton subtraction}
\label{sec: prelims-PS}
 Unless otherwise mentioned a beamsplitter of transmittance $T$ (reflectance $R$) mixes incident annihilation operators in a Bogoliubov transform given by $\hat{a}_1\mapsto\sqrt{T}\hat{a}_1-\sqrt{R}\hat{a}_2$ \mbox{and $\hat{a}_2\mapsto \sqrt{R}\hat{a}_1+\sqrt{T}\hat{a}_2$.} A general $n$-photon subtraction in optical systems is performed by splitting the beam and heralding the output in the transmitted signal on the detection of $n$ photons in the reflected~ancilla. 
 
 It can be easily seen that when $n$ photons are detected, it implements an operation $\propto \sqrt{T}^{\hat{n}}\hat{a}^n$ where $\hat{a}$ and $\hat{n}$ are the photon-annihilation and number \mbox{operators, respectively.} This is the same as noiseless attenuation when $n=0$ photons are detected \cite{mivcuda2012noiseless, nunn2021heralding} for any input state, $\ket{\psi}=\sum_{n=0}^{\infty}c_n\ket{n}$ as $T<1$. Also note that the same $\sqrt{T}^{\hat{n}}$ that is responsible for noiseless attenuation remains when~$n>0$.

\subsection{Teleamplification}
\label{sec: prelims-teleamp}

Standard teleamplification implements the operation $\propto g^{\hat{n}}$ just as photon subtraction implements noiseless attenuation. However, it only works for inputs that are superpositions of coherent states having the same absolute value of the amplitude, i.e. they lie on a ring. These states are optical cat states.

Teleamplification works by heralding the output state on the detection of several photons in various ancillas. In general, it implements $\propto g^{\hat{n}}f(\hat{a})$ where $f(\hat{a})$ is some function of the annihilation operators. The input is a superposition that can be written as $\ket{\psi}=\sum_kc_k\ket{\alpha u^k}$ for fixed magnitude $\alpha$ and phase $u$ \cite{neergaard2013quantum} for arbitrary coefficients $c_k$. Note that when inputs are restricted to this form, operations $f(\hat{a})$ and $g^{\hat{n}}$ commute to an irrelevant global factor that depends on the magnitude $\alpha$, because coherent states are eigenstates of the photon-annihilation~operator.

When $n$-photon subtraction and teleamplification are put together (for such input cat states), we get the operation $\propto \hat{a}^{n}f(\hat{a})(\sqrt{T}g)^{\hat{n}}$. If $g=1/\sqrt{T}$ we get an operation that does not change the magnitude of the coherent state in the superposition.

\subsection{Photon loss}
\label{sec: prelims-loss}

In a typical communication channel such as a fiber optic cable of length $l$ and attenuation length $L_{{\rm{attn.}}}$, the noise experienced during travel time $t$ is modeled, in the rotating frame, by the LGKS master equation \cite{lindblad1976generators, gorini1976completely}
$\partial_t \hat\rho=\kappa\mathcal{D}[\hat{a}](\hat{\rho})
$ with the dissipation rate $\kappa$ and the \mbox{dissipator} $\mathcal{D}[\hat{A}]=\hat{A}\bullet\hat{A}^\dagger-\{\hat{A}^\dagger\hat{A},\bullet\}/2$ for a jump operator $\hat{A}$.

The solution to this is the loss or the amplitude-damping quantum channel \cite{chuang1997bosonic} with a Kraus representation \cite{kraus1983states}, 
\mbox{$\mathcal{E}_\gamma (\hat{\rho})=\sum_{k=0}^{\infty}\hat{E}_k\hat\rho\hat{E}_k^\dagger,$}
where \mbox{$\hat{E}_k = \sqrt{\left( 1 - \gamma \right)^k/k!} \sqrt{\gamma}^{\hat{n}} \hat{a}^k$} 
with the channel transmittance, $\gamma = e^{-\kappa t}=e^{-l/L_{{\rm{attn.}}}}$.
The set of operators $\{\hat{E}_k^\dagger\hat{E}_k\}$ forms a POVM, with the completeness or the trace-preserving condition, $\sum_{k=0}^{\infty}{\hat{E}_k^\dagger\hat{E}_k}=\hat{I}$ \cite{nielsen_chuang_2010}.

\subsection{Multicomponent cat codes}
\label{sec: prelims-cat-codes}
Even and odd coherent states are defined as \cite{dodonov1974even} 
\begin{equation}
    \smlket{\overline{0}} = \mathcal{N}_{+}\argp{\ket{\alpha} + \ket{-\alpha}}\ {\rm{and}}\ \smlket{\overline{1}} = \mathcal{N}_{-}\argp{\ket{\alpha} - \ket{-\alpha}}
\end{equation}
where the normalization is $
    \mathcal{N}_{\pm}=1/\sqrt{2\argp{1\pm e^{-2\alpha^2}}}$, the overline denoting the logical basis, and assuming real $\alpha$. These are superpositions of coherent states that are symmetrically spaced on a ring in the phase space and only of even and odd Fock states. These states are now commonly known as the two-component cat codes. 
    
    They can be used for efficient quantum computation \cite{jeong2002efficient,ralph2003quantum} and generated using single-photon sources and simple all-optical operations \cite{lund2004conditional,ourjoumtsev2007generation} or weak nonlinearity \cite{jeong2004generation}. Although they cannot correct logicl errors due to photon-losses by themseleves, they are shown to be useful in increasing the fault tolerance threshold in the biased noise regime in some platforms where dephasing errors are more likely than relaxation errors \cite{aliferis2009fault, tuckett2018ultrahigh}. When an odd number of photons is annihilated, it flips their photon-number parity, whereas when the annihilation of an even number of photons leaves it unchanged.
    
Another set of cat codes, the four-component cat codes \cite{mirrahimi2014dynamically} given by 
\begin{equation}
\label{eq: 4_cat}
    \ket{\overline{\mu}} = \mathcal{N}_+\argp{\ket{\I^\mu \alpha}+\ket{- \I^\mu \alpha}},
\end{equation}
with $\mu=0,1$, have recently achieved break-even with the bare encoding in \mbox{circuit-QED} \cite{ofek2016extending}.  These codes are resilient to single-photon losses, as they can be detected and corrected with this~code. Single-photon annihilation on these code states produce 
\begin{equation}
    \ket{\overline{\mu}'} = \mathcal{N}_-\argp{\ket{\I^\mu \alpha}-\ket{- \I^\mu \alpha}},
\end{equation}
which have odd parity.

A qubit may be encoded on the basis given by Eq. (\ref{eq: 4_cat}), or alternatively on 
\begin{equation}
\smlket{\overline{\pm}}={\mathcal{M}_\pm}\args{\ket{\alpha}+\ket{-\alpha}\pm \argp{\ket{\I\alpha}+\ket{-\I\alpha} }}
\label{eq: plus minus basis}
\end{equation}
with ${\mathcal{M}}_{\pm} = {1}/{\args{2\sqrt{2e^{-{\alpha}^2}\argp{\cosh{\alpha^2}\pm \cos{\alpha^2}}}}},$

The corresponding odd parity states that form yet another orthonormal basis are
\begin{equation}
    \smlket{\overline{\pm}'} = \mathcal{M}^{'}_{\pm}\args{\ket{\alpha}-\ket{-\alpha}\pm \I \argp{\ket{\I\alpha}-\ket{-\I\alpha} }}
\end{equation}
with $\mathcal{M}^{'}_{\pm} = {1}/{\args{2\sqrt{2e^{-{\alpha}^2}\argp{\sinh{\alpha^2}\mp \sin{\alpha^2}}}}}.$

\section{Error suppression on encoded states}
\label{sec: error suppression}

\subsection{With perfect detectors}
Consider four-component cat code given by Eq. (\ref{eq: 4_cat}) and a qubit encoded on them as \mbox{$\smlket{\overline{\psi}} = \sum_{k=0}^3c_k\ket{\alpha_k}$} defining $\alpha_k = \alpha u^k$ with $u=e^{\frac{2\pi\I}{4}}=\I$, where we note that $u$ takes on a general phase factor for other multicomponent cat codes \cite{neergaard2013quantum}. An ideal $n$-photon subtraction with a beamsplitter of transmittance $T$ effectively implements an operation
$\propto \sqrt{T}^{\hat{a}^\dagger\hat{a}}\hat{a}^{n}$. When~$n=0$, it is the same as noiseless attenuation. 

An arbitrary encoded state $\smlket{\overline{\psi}} \propto {a\smlket{\overline{0}}+b\smlket{\overline{1}}}.
$
where the logical 0 and 1 states are two-component even cats having identical normalizations, $\mathcal{N}_+$, so the state is proportional to
${a\argp{\ket{\alpha}+\ket{-\alpha}}+b\argp{\ket{\I\alpha}+\ket{-\I\alpha}}}.
$
As an example, after an ideal zero-photon subtraction (ZPS) with beamsplitter of transmissivity $\sqrt{T}$, the coherent-state amplitudes are noiselessly attenuated by the same factor giving the normalized state
\mbox{$\smlket{\overline{\psi}}  =\frac{a\smlket{\widetilde{0}}+b\smlket{\widetilde{1}}}{\sqrt{1+2{\rm Re}\argp{a^*\smlbraket{\widetilde{0}}{\widetilde{1}}b}}},
$}
where all quantities with a tilde correspond to \mbox{reduced amplitudes.} This state is heralded by the detection of zero photons with a \mbox{success probability,}
\begin{align}
P_{\rm{s}} = \argp{\frac{\mathcal{N}_+}{\widetilde{\mathcal{N}}_+}}^2 \frac{1+2{\rm Re}\argp{a^*\smlbraket{\widetilde{0}}{\widetilde{1}}b}}{1+2{\rm Re}\argp{a^*\smlbraket{\overline{0}}{\overline{1}}b}}e^{-R\alpha^2}.
\end{align}

\begin{figure}[htbp]
\centering
\includegraphics[width=10cm]{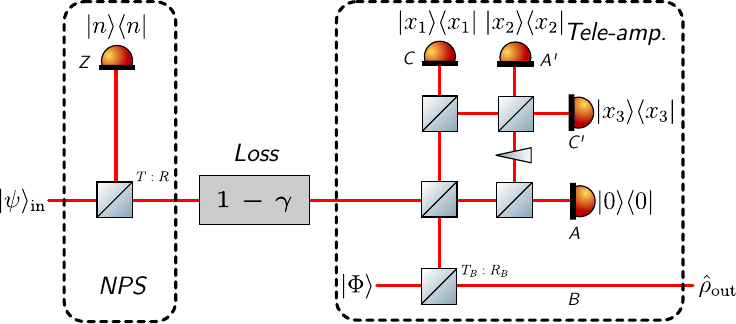}
\caption{ Setup for error suppression in four-component \mbox{cat codes.} Here, zero-photon subtraction implements noiseless attenuation and the standard teleamplification implements  noiseless amplification. The standard teleamplification requires $|\Phi\rangle$ a resource of a particular large-amplitude cat state and coincident detection of single photons. Nonzero photon subtraction and a more general heralding for teleamplification is also considered with all $x_k\neq0$. A vacuum state is heralded along with the desired state in one of the outputs, here it is the mode $A$. The unmarked beamsplitters are balanced $50:50$, and the triangle corresponds to the phase \mbox{shift by $\pi/2$.} The output from the amplification is sent for recovery by \mbox{error correction.} }
\label{fig: setup}
\end{figure}

\begin{figure}[htbp]
\centering
\includegraphics[width=10cm]{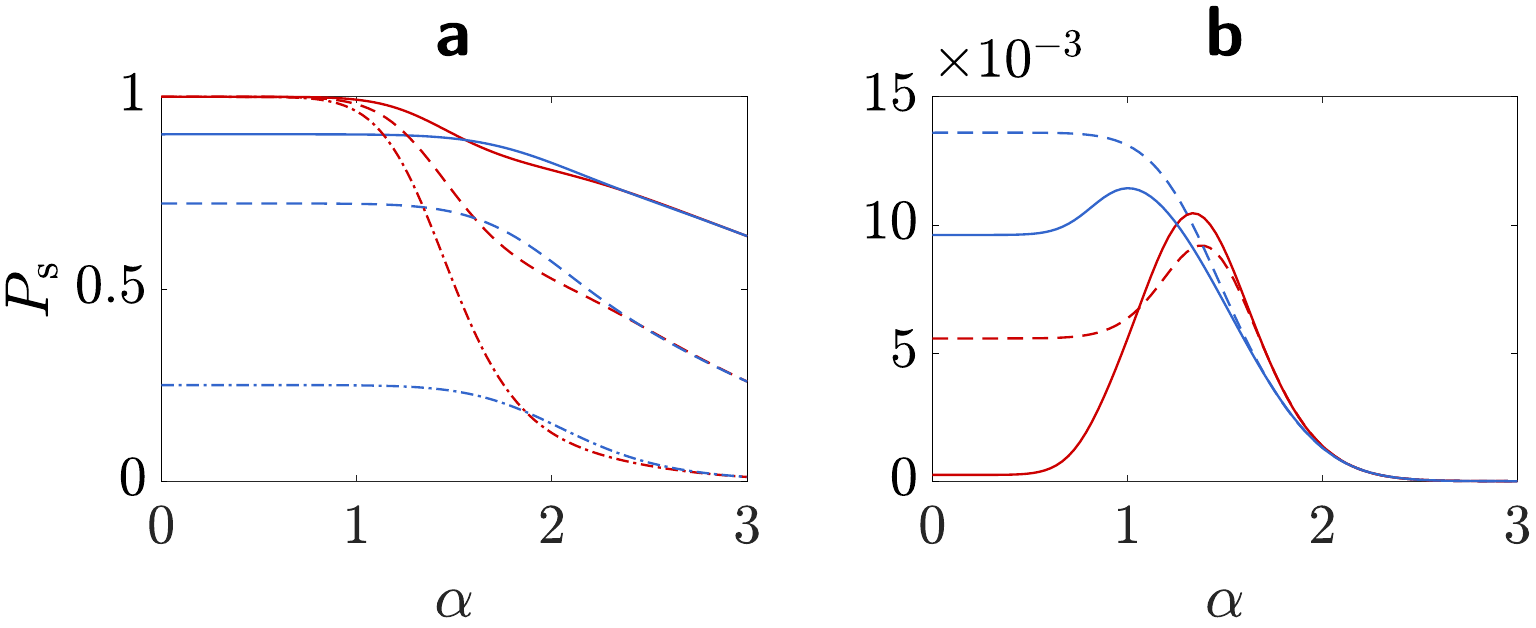}
\caption{The success probability versus input coherent-state amplitude for (a) standalone ZPS, with beamsplitter transmittances: $T=95\%$ (solid), $85\%$ (dashed), and $50\%$ (dot-dashed), and for (b) standalone standard teleamplification ($x_j=1$) with gains: $g=1.25$ (dashed) and $g=2.5$ (solid), for the inputs with $\smlket{\overline{+}}$ (red) and $\smlket{\overline{\vphantom{+}-}}$ (blue). Perfect detectors have been assumed. }
\label{fig: Ps}
\end{figure}

A teleportation-based approach \cite{neergaard2013quantum} is employed to noiselessly amplify the coherent-state amplitudes as shown by Fig. \ref{fig: setup}. A resource cat state with a large coherent-state amplitude, say $\beta$, is required. This state is chosen as 
$
\ket{\Phi}\propto\smlket{\overline{0}'}+\I\smlket{\overline{1}'}
=\mathcal{M}_{+}^{'}\sum_{k=0}^{3}{u^{-3k}\ket{\beta_k}},
$
where $\beta_k=u^{k} \beta$ and $\mathcal{M}_{+}^{'}$ is a function of the amplitude $\beta$.  If $\beta = \alpha/\sqrt{R_B}$ and
$g = \sqrt{T_B/R_B}$ then the output state is given by \cite{neergaard2013quantum}
\begin{align}
\begin{split}
\ket{\psi}=\mathcal{M}_{+}^{'} \sum_{j,k=0}^{3}{c_j} u^{-3k} &\ket{g \alpha_k}_B 
\ket{\frac{\alpha}{2}(u^j-u^k)}_A\ket{\frac{\alpha}{2\sqrt{2}} \argp{u^j-u^{j+1} +u^k +u^{k+1}} }_{A'}
\\
&\ket{-\frac{\alpha}{2}(u^j+u^k)}_C\ket{\frac{\alpha}{2\sqrt{2}} \argp{u^j+u^{j+1} +u^k -u^{k+1}} }_{C'}.
\end{split}
\label{eq: ideal teleamp}
\end{align}
It can be verified that for any given pair $(j,k)$ there is one mode guaranteed to be in the vacuum. 

In our case, this is the mode $A$ whenever $j=k$. Therefore, conditionally measuring strictly-nonzero Fock states $(x_1, x_2, x_3)$ coincident in modes $A'$, $C$, and $C'$ heralds a vacuum \mbox{in mode $A$.} 

Assuming a real-valued $\alpha$ under such a heralding, we get the state,
\begin{equation}
\mathcal{M}_{+}^{'} \frac{e^{-\alpha^2}\alpha^{x_1+x_2+x_3}}{\sqrt{x_1!x_2!x_3!2^{x_2+x_3}}}\sum_{k=0}^{3}{c_k u^{k(x_1+x_2+x_3-3)}\ket{g \alpha_k}}_B.
\end{equation}
The success probability of heralding the desired state is given by
\begin{equation}
    P_{\rm{s}} = \argp{\mathcal{M}_{+}^{'} \frac{e^{-\alpha^2}\alpha^{x_1+x_2+x_3}}{\sqrt{x_1!x_2!x_3!2^{x_2+x_3}}}}^2
    \sum_{j,k=0}^{3}{c_j^*u^{(-j+k)(x_1+x_2+x_3-3)}c_k e^{-g^2\alpha^2\argp{ 1- u^{-j+k} }}}.
\end{equation}
An ideal teleamplification thus implements the operation $\propto \hat{a}^{x_1+x_2+x_3-3}g^{\hat{a}^\dagger \hat{a}}$ where $(x_1,x_2,x_3)$ are the numbers of (nonzero) photons detected in modes $C$, $A'$ and $C'$ as shown in Fig. \ref{fig: setup}. This is valid only because the input is restricted to cat states where the magnitudes of coherent-state amplitudes are all identical and they have a discrete rotation symmetry~\cite{grimsmo2020quantum}. It typically does not hold for any arbitrary state in the phase space. When $(x_1,x_2,x_3)=(1,1,1)$ we get back the standard teleamplification scheme from Ref. \cite{neergaard2013quantum}.

The combined operation of $n$-photon subtraction followed by general teleamplification, in the absence of any noise and when $g=1/\sqrt{T}$, is $\propto \hat{a}^{(n+x_1+x_2+x_3-3)}$. This is simply a subtraction of the $(n+x_1+x_2+x_3-3)$ number of photons from the input cat states. 

\begin{figure}[h]
\centering
\includegraphics[width=9cm]{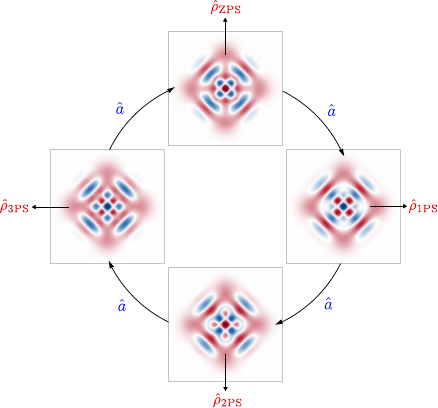}
\caption{The result of multiphoton subtraction and the general teleamplification together on arbitrary encoded states give a finite number of possible outcomes and corresponds to an effective $n'({\rm mod}\,4)$ photon subtraction. The Wigner distributions of $\smlket{\overline{+}}=\mathcal{M}_+\sum_{k=0}^3\ket{\I^k\alpha}$ under different numbers of subtracted photons are shown here. It cycles between 4 outcomes due to the discrete rotation symmetry of the code along with the coherent state being the eigenstates of the \mbox{photon-annihilation operator.}
The final state is one of the four $\hat{\rho}_{n'({\rm {mod}}\ 4) {\rm PS}}$ mixtures when losses are considered. Error correction procedure is adapted to each of these outcomes separately.}
\label{fig: encoding-symmetry}
\end{figure}

Any arbitrary number of photon subtractions leads to a finite number of possible outcomes when the qubit is encoded on cat states. We have $
\hat{a}^{n'}\mapsto \hat{a}^{n'({\rm mod}\ 4)}$ for any $n'$ for four-component cats. Therefore, the final state is only one of four possible mixtures, those corresponding to $n'=n+x_1+x_2+x_3-3=0,\,1,\,2,\,{\rm and}\ 3~({\rm mod}~4)$ subtracted photons rather than infinitely many. All combinations of detected photon numbers $(n,x_1,x_2,x_3)$ such that $x_j\neq 0$ herald the same mixture $\hat{\rho}_{n'({\rm mod}\,4){\rm PS}}$, as depicted by Fig. \ref{fig: encoding-symmetry}.

An amplitude-damping channel acting on the state $\propto a\smlket{\overline{0}}+b\smlket{\overline{1}}$ modified by noiseless attenuation by a factor $\sqrt{T}$ gives the mixture (using Eq. (13) of Ref. \cite{bergmann2016quantum}), 
\begin{align}
\begin{split}
    \widetilde{\rho} = &\,\widetilde{p}_0 N_0^{(\sqrt{\gamma T }\alpha)}
    \argp{a\smlket{\widetilde{0}}+b\smlket{\widetilde{1}}}\times {\rm{H.c.}}
    +\widetilde{p}_1 N_1^{(\sqrt{\gamma T } \alpha)}
    \argp{a|\widetilde{0}'\rangle+\I b|\widetilde{1}'\rangle}\times {\rm{H.c.}}
    \\
    + &\,\widetilde{p}_2 N_2^{(\sqrt{\gamma T } \alpha)}
    \argp{a\smlket{\widetilde{0}}-b\smlket{\widetilde{1}}}\times {\rm{H.c.}}
    +\widetilde{p}_3 N_3^{(\sqrt{\gamma T } \alpha)}
    \argp{a|\widetilde{0}'\rangle-\I b|\widetilde{1}'\rangle}\times {\rm{H.c.}}
    \label{eq: rho}
\end{split}
\end{align}
where the states with tilde correspond to the final amplitude $\sqrt{\gamma T}\alpha$ after the amplitude damping channel and we have defined the normalizations,
\begin{align}
    \begin{split}
        N_0^{(\alpha)}&=\frac{1}{\sqrt{1+2{\rm Re}\argp{ab^*} {\cos{\argp{ \alpha^2}}}/{\cosh{\argp{\alpha^2}}}}},
        N_1^{(\alpha)}=\frac{1}{\sqrt{1-2{\rm Re}\argp{ab^*} {\sin{\argp{ \alpha^2}}}/{\sinh{\argp{\alpha^2}}}}},
        \\
        N_2^{(\alpha)}&=\frac{1}{\sqrt{1-2{\rm Re}\argp{ab^*} {\cos{\argp{ \alpha^2}}}/{\cosh{\argp{\alpha^2}}}}},
        N_3^{(\alpha)}=\frac{1}{\sqrt{1+2{\rm Re}\argp{ab^*} {\sin{\argp{ \alpha^2}}}/{\sinh{\argp{\alpha^2}}}}}.
    \end{split}
\end{align}
The weights of the mixture are given by 
\begin{equation}
   \widetilde{p}_k = p_k N_0^{(\sqrt{T}\alpha)}/N_k^{(\sqrt{\gamma T}\alpha)} 
\end{equation}
for $k=0,1,2,3$, where
\begin{align}
\label{eq: loss_prob}
\begin{split}
    p_0 = {\cosh\argp{\gamma T \alpha^2}} 
    &\left\{\cosh\args{\argp{1-\gamma}T\alpha^2}+\cos\args{\argp{1-\gamma}T\alpha^2}\right\}/{2\cosh{\argp{T\alpha^2}}},
    \\
    p_1 =  {\sinh\argp{\gamma T \alpha^2}} 
    &\left\{\sinh\args{\argp{1-\gamma}T\alpha^2} +\sin\args{\argp{1-\gamma}T\alpha^2}\right\}/{2\cosh{\argp{T\alpha^2}}},
    \\
    p_2 = {\cosh\argp{\gamma T \alpha^2}} 
    &\left\{\cosh\args{\argp{1-\gamma}T\alpha^2}-\cos\args{\argp{1-\gamma}T\alpha^2}\right\}/{2\cosh{\argp{T\alpha^2}}},
    \\
    p_3 = {\sinh\argp{\gamma T \alpha^2}} 
    &\left\{\sinh\args{\argp{1-\gamma}T\alpha^2}-\cos\args{\argp{1-\gamma}T\alpha^2}\right\}/{2\cosh{\argp{T\alpha^2}}}.
\end{split}
\end{align}

To recover the quantum information by error correction, the following map is provided by~Ref.~\cite{bergmann2016quantum}. The states associated with even photon number parity, those corresponding to the probabilities $\widetilde{p}_0$ and $\widetilde{p}_2$, are mapped to the codespace as is, identically, and the states associated with odd photon number parity, those corresponding to the probabilities $\widetilde{p}_1$ and $\widetilde{p}_3$, are mapped to the codespace after a parity flip and a phase rotation of $\pi/2$. The worst-case fidelity over all input states was shown to be $\mathcal{F}_{\rm w} = \min_{|a|,|b|=1/\sqrt{2}}(\widetilde{p}_0+\widetilde{p}_1)$ \cite{bergmann2016quantum}. However, this mapping defines a particular recovery, which may generally be suboptimal. Optimal recoveries are discussed \mbox{in Sec. \ref{sec: optimal-recovery}.}

\subsection{With imperfect detections}
Detector efficiency plays a significant role in any \mbox{conditional measurement.} In practice, superconducting-nanowire single-photon detectors with system detection efficiency $>98\%$ and a small dark count rate of $10^{-4}$ Hz have been realized \cite{esmaeil2021superconducting}. More ubiquitous single-photon avalanche diodes offer modest detection with efficiencies $\sim50\%$, for reasonably low dark count rates \mbox{of $\sim10^2$ Hz~\cite{restelli2013single, nunn2021heralding}.} In all of our discussions in this paper, we assume a negligible dark count rate for \mbox{convenience and simplicity.} 

The composition of encoding ($\mathcal{C}$), photon subtraction ($\mathcal{Z}$), loss ($\mathcal{E}_\gamma$), teleamplification ($\mathcal{T}$), recovery ($\mathcal{R}$), and decoding ($\mathcal{C}^{-1}$) is the channel 
\begin{equation}
    \mathcal{E}=\mathcal{C}^{-1}\circ\mathcal{R}\circ\argp{\mathcal{T} \circ \mathcal{E}_\gamma \circ \mathcal{Z}} \circ \mathcal{C}.
\end{equation}
Assume that the encoding, recovery, and decoding are ideal and are implemented instantaneously.

 To simplify the analysis, the combined loss suppression setup can be recast into an equivalent one as follows: (1)  noiseless amplification by a factor $T_{\rm{eff.}}$ followed by (2) ideal subtraction of $(n+x_1+x_2+x_3-3)$ photons, and finally (3) effective loss $(1-\gamma_{\rm{eff.}})$. The losses arising from detector inefficiencies and the loss channel itself are combined into an effective loss channel ($\mathcal{E}_{\gamma_{\rm {eff.}}}$) and, similarly, the attenuation that comes from photon subtraction and amplification from the teleamplification is combined into an effective noiseless rescaling ($\mathcal{T_{\rm{eff.}}}$) so that we have
\begin{equation}
    \mathcal{E}=\mathcal{C}^{-1}\circ\mathcal{R}\circ\argp{ \mathcal{E}_{\gamma_{\rm {eff.}}} \circ \mathcal{T_{\rm{eff.}}} }\circ \mathcal{C}.
    \label{eq: effective-channel}
\end{equation}
The effective channels are described by the parameters given in Eq. (\ref{eq: effective T and gamma}). This leads to the same state as the original scheme (Appendix \ref{append_sec: equivalent-setups}). However, the success probabilities of heralding in the original setup cannot be calculated with this equivalent setup.

\section{Optimizing the recovery channel}
\label{sec: optimal-recovery}

Here we present the procedure for optimizing the worst-case-fidelities of the recovered qubit over completely-positive trace-preserving (CPTP) recovery maps. Instead of encoding on the $(\overline{0},\overline{1})$ basis presented in Eq. (\ref{eq: 4_cat}) it is more convenient to use the physical states $(\overline{+},\overline{\vphantom{+}-})$ to encode the states of the computational basis $\ket{\mu_{\rm{comp.}}}$ using a channel $\mathcal{C}=C\bullet C^{\dagger}$
where 
$
    C = \ketbra{\overline{+}_{\rm{cat}}}{0_{\rm{comp.}}}+\ketbra{\overline{\vphantom{+}-}_{\rm{cat}}}{1_{\rm{comp.}}}.
$
The subscripts will be suppressed going forward.

The states $\smlket{\overline{0}}$ and $\smlket{\overline{1}}$ and similarly $\smlket{\overline{0}'}$ and $\smlket{\overline{1}'}$ are not orthogonal to each other, while the states $\propto \smlket{\overline{0}}\pm\smlket{\overline{1}}$ and $\propto \smlket{\overline{0}'}\pm\smlket{\overline{1}'}$ are all orthogonal, making them attractive choices for a basis. 
However, the qubits encoded on these suffer from distortions under noiseless attenuation or amplification. Consider the encoded state $
\smlket{\overline{\psi}}=a\smlket{\overline{+}}+b\smlket{\overline{\vphantom{+}-}}$
which, after noiseless attenuation (with ZPS) or amplification, turns into  
\begin{equation}
\propto a\frac{\mathcal{M}_+}{\widetilde{\mathcal{M}}_+}\smlket{\widetilde{+}}+b\frac{\mathcal{M}_-}{\widetilde{\mathcal{M}}_-}\smlket{\widetilde{\vphantom{+}-}},
\label{eq: distorted-qubit}
\end{equation}
where the quantities with tilde correspond to the modified coherent-state amplitudes.

This shows a redistribution of the coefficients that distorts the qubits. The final state is pure, assuming that all operations are ideal, and simply differs from the original one by a rotation on the Bloch sphere. If the qubit is perfectly known initially, then this distortion can be undone. However, for an unknown arbitrary input qubit, this is a nonunitary transformation.

This problem is not unique to codes modified by noiseless attenuation.  When the encoded qubits pass through some lossy channel, their coherent-state amplitudes decrease, also giving rise to this problem. This needs to be addressed by a suitable amplification that compensates for this distortion, as explained in Sec. \ref{sec: prelims-teleamp}.

Now we describe constructing the basis operators and the corresponding process and Choi matrices for the case of $n'(\textrm{mod}\,4)=0, 2$, which are effective ZPS and 2PS, respectively, during the operation $\mathcal{T}_{\rm eff.}$. Similar constructions are carried out for the other \mbox{two cases.}

\subsection{Orthonormal basis operators for the noise and recovery channels}
\label{sec: basis operators}

The choice of states of $(\overline{+},\overline{\vphantom{+}-})$ as a basis can be used to construct an orthonormal set of basis operators $\{\hat{B}_i\}$ for the Kraus operators of the recovery channel as
\begin{alignat}{8}
        &\hat B_{0e} = \Big(&&\smlketbra{\overline{+}}{\widetilde{+}}&& + &&\smlketbra{\overline{\vphantom{+}-}}{\widetilde{\vphantom{+}-}} ~\Big)\Big/{\sqrt 2}\,,\,
       &&\hat B_{1e} = \Big(&&\smlketbra{\overline{\vphantom{+}-}}{\widetilde{+}}&& + &&\smlketbra{\overline{+}}{\widetilde{\vphantom{+}-}}~\Big)\Big/{\sqrt 2}\,, 
       \nonumber
       \\
       &\hat B_{2e} = \I\Big(&&\smlketbra{\overline{\vphantom{+}-}}{\widetilde{+}}&& - &&\smlketbra{\overline{+}}{\widetilde{\vphantom{+}-}} ~\Big)\Big/\sqrt 2\,,\quad
       &&\hat B_{3e} = \Big(&&\smlketbra{\overline{+}}{\widetilde{+}}&& - &&\smlketbra{\overline{\vphantom{+}-}}{\widetilde{\vphantom{+}-}} ~\Big)\Big/\sqrt{2}\,,
\end{alignat}
that project states in the errorspace formed as a consequence of even-numbered photon annihilations after the effective loss channel to the codespace and similarly defined operators $\{\hat{B}_{jo}\}$ using the states $\smlket{\widetilde{\pm}'}$, rather than $\smlket{\widetilde{\pm}}$, do the same for the states after odd-numbered \mbox{photon annihilations.} 

The quantities with tilde represent the coherent-state amplitude just after the effective loss channel, whereas those with a bar represent the restored amplitudes. The basis operators for the loss channel can have the same form as the Hermitian conjugates to these operators, but with the tilde representing the coherent-state amplitude after the loss channel and the bar representing the amplitude after effective noiseless rescaling, $\mathcal{T}_{\rm eff.}$. 

With this operator basis, we can now create the process \cite{poyatos1997complete} and Choi \cite{choi1975completely} matrices of the loss and \mbox{recovery channels.}

\subsection{Process and Choi matrices}
\label{sec: process matrix}

Let ${\hat{R}_r}$ and ${\hat{E}_k}$ be the Kraus operators of the recovery and effective loss channels respectively. The recovery process matrix is defined by elements $X_{ij}= \sum_{r} {\chi_{ri} \chi^*_{rj}}$ such that $\hat{R}_r= \sum_l \chi_{rl} \hat{B}_l$ for the set of basis operators $\{\hat{B}_l\}$, and the process matrix for the effective loss channel is defined by elements $W_{ij}= \sum_{k} \omega_{ki} \omega_{kj}^* $ such \mbox{that $\hat{E}_k = \sum_l \omega_{kl} \hat{B}_l^\dagger$.} Both $X$ and $W$ are Hermitian matrices and the index $k$ runs over all the Kraus operators of the channel, corresponding to different numbers of photon losses, whereas the index $i$ runs over all the \mbox{basis operators.}

 We can conveniently arrange all the terms, $\omega_{ki}$, into different matrices of a block-diagonal form \mbox{$w_n = {\rm diag}(\omega_e, \omega_o),$}
for \mbox{$n=0,1,2,\dots$}, such that the rows correspond to only four Kraus operators for any given $n$, $\{\hat{E}_{4n}, \hat{E}_{4n+2}, \hat{E}_{4n+1}, \hat{E}_{4n+3}\}$, 
and the columns correspond to all the basis operators, $\{\hat{B}_i^\dagger\}$, forming the matrix elements, $[w_n]_{ki} = {\rm tr}\{\hat{E}_k \hat{B}_i\}$.

Such a construction gives
\begin{align}
    &\omega_e = \begin{pmatrix}
         \omega_{4n,0e} & 0 & 0 & \omega_{4n, 3e} \\
        0 & \omega_{4n+2, 1e} & \omega_{4n+2, 2e} &0
    \end{pmatrix},
\end{align}
with
   \begin{align}
   \begin{split}
      &\omega_{4n, 0e} = \frac{1}{\sqrt{2}}\smlbraket{4n}{\sqrt{1-\gamma}\alpha} \argp{\frac{\mathcal{M}_+}{\widetilde{\mathcal{M}}_+}+\frac{\mathcal{M}_-}{\widetilde{\mathcal{M}}_-} },\,
       \omega_{4n+2, 1e} = \frac{1}{\sqrt{2}}\smlbraket{4n+2}{\sqrt{1-\gamma}\alpha} \argp{\frac{\mathcal{M}_-}{\widetilde{\mathcal{M}}_+}+\frac{\mathcal{M}_+}{\widetilde{\mathcal{M}}_-} },
      \\
      &\omega_{4n, 3e} = \frac{1}{\sqrt{2}}\smlbraket{4n}{\sqrt{1-\gamma}\alpha}  \argp{\frac{\mathcal{M}_+}{\widetilde{\mathcal{M}}_+}-\frac{\mathcal{M}_-}{\widetilde{\mathcal{M}}_-} },\,
      \omega_{4n+2, 2e} = \frac{1}{\sqrt{2}}\smlbraket{4n+2}{\sqrt{1-\gamma}\alpha}  \argp{\I\frac{\mathcal{M}_-}{\widetilde{\mathcal{M}}_+}-\I\frac{\mathcal{M}_+}{\widetilde{\mathcal{M}}_-} }.
   \end{split}
   \end{align}
The matrix
$\omega_o$ is similarly 
\begin{align}
    &\omega_o = \begin{pmatrix}
         \omega_{4n+1,0o} &  0 &  0 & \omega_{4n+1, 3o} \\
        0 & \omega_{4n+3, 1o} & \omega_{4n+3, 2o} & 0
    \end{pmatrix},
\end{align}
with $\omega_{k, jo}$ containing $\widetilde{\mathcal{M}}^{'}_{\pm}$ rather than $\widetilde{\mathcal{M}}_{\pm}$ in \mbox{the denominators.}

The process matrix is given by the elements 
\begin{align}
       W_{ij} = \sum_{k=0}^{\infty} {\omega_{k, i} \omega^*_{k,j}} = \argp{\sum_{n=0}^{\infty}w^\intercal_n w^*_n}_{ij}= \argp{\sum_{n=0}^{\infty}w^\dagger_n w_n}_{ji} \cdotp
   \end{align}
   We may truncate the sum to some reasonable value as an approximation while using the Fock basis. However, the current problem has a closed analytical form. The accompanying \textsc{mathematica} files contain the expressions.

The process matrix $X$ can be easily transformed into the corresponding Choi matrix representation $J_X$ as\cite{teo2013informationally},
\begin{equation}
    J_X = \sum_{kk'}\sket{{B}_k}X_{kk'}\sbra{{B}_{k'}},
\end{equation}
where $|\bullet\rangle\mkern-5mu\rangle$ (or ${\rm {vec}}[\bullet]$) is the vectorization of matrices \cite{teo2015introduction, gyamfi2020fundamentals} in the computational basis with the column-stacking convention being used here. The Choi matrix $J_W$ is constructed in a similar manner.

\subsection{The algorithm}
\label{sec: algorithm}

We now outline a gradient ascent algorithm to find the optimal worst-case fidelity, 
\begin{equation}
    \mathcal{F}_{\rm w}^{\rm max} = \max_{\mathcal{R}} \min_{\ket{\psi}}  {\rm {tr}}\argc{ \smlketbra{\psi}{\psi} \mathcal{E}\args{\smlketbra{\psi}{\psi}}},
\end{equation}
and the corresponding recovery with the tools prepared in the previous sections as follows.

First, the optimal Choi matrix, $\mathcal{R}$, is initiated as a random CPTP matrix (Appendix \ref{append_sec: CPTP init}) and then:
\begin{enumerate}
\item The pure state $\ket{\psi}$ that minimizes the fidelity is found. In \textsc{matlab}, this is easily done using the \texttt{fmincon} function.
\item The gradient in $\mathcal{R}$ is calculated for the \mbox{state $\ket{\psi}$ (Appendix \ref{append_sec: grad-fidelity}).}
\item A small CPTP-constrained step $\delta\mathcal{R}$ (Appendix \ref{append_Sec: CPTP-deltaR}) is taken in the direction of the steepest gradient ascent (Appendix \ref{append_sec: deltaA}). 
\end{enumerate}
These steps are repeated with a decreasing stepsize until the fidelity becomes constant up to some tolerance or for a sufficiently large number of steps. 

The domain is the surface of the Bloch sphere, which makes the nonconvex and precludes the use of semi-defintie programs \cite{albert2018performance, schlegel2022quantum} in this context. The nonconvex optimization for a global optimal value is performed by repeating the process with different initial conditions and taking the best values among the results.

\section{Results}
\label{sec: results}

This section compiles the results from the methods discussed in the previous sections, while highlighting the key findings of the work.

\begin{figure}[h]
\centering
\includegraphics[width=10cm]{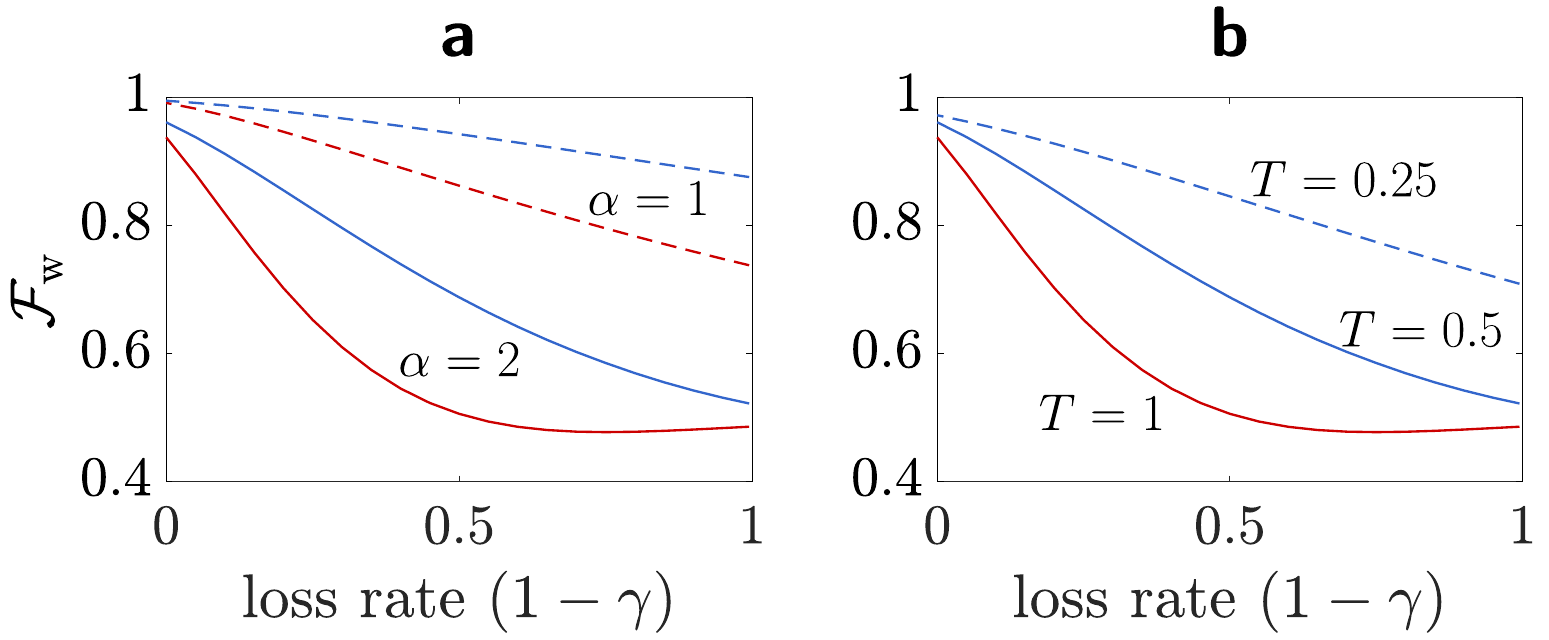}
\caption{The performance of four-component cat code through a loss channel followed by amplitude restoration by realistic, noisy teleamplification, with (blue) and without (red) \mbox{noisy ZPS.} The back action of ZPS suppresses uncorrectable errors in the encoded qubit for a wide range of parameters. For each graph, only one parameter is varied while the fixed parameters take the values: $\alpha=2$, $T=50\%$. The detection efficiency is fixed at $\eta=95\%$.}
\label{fig: full-setup-effectiveness}
\end{figure}

The worst-case fidelity for the qubits encoded on $(\overline{0}, \overline{1})$ basis, $\mathcal{F}_{\rm w}$ due to Ref. \cite{bergmann2016quantum}, which is modified under loss suppression is compared to the corresponding worst-case fidelity under teleamplification alone in Fig. \ref{fig: full-setup-effectiveness}. It clearly illustrates the benefit of the noiseless loss suppression scheme with ZPS and standard teleamplification as a \mbox{viable} method of error suppression for various system parameters and across the different values of the loss rates for a realistic value of detector efficiency. 

\begin{figure}[t]
\centering
\includegraphics[width=10cm]{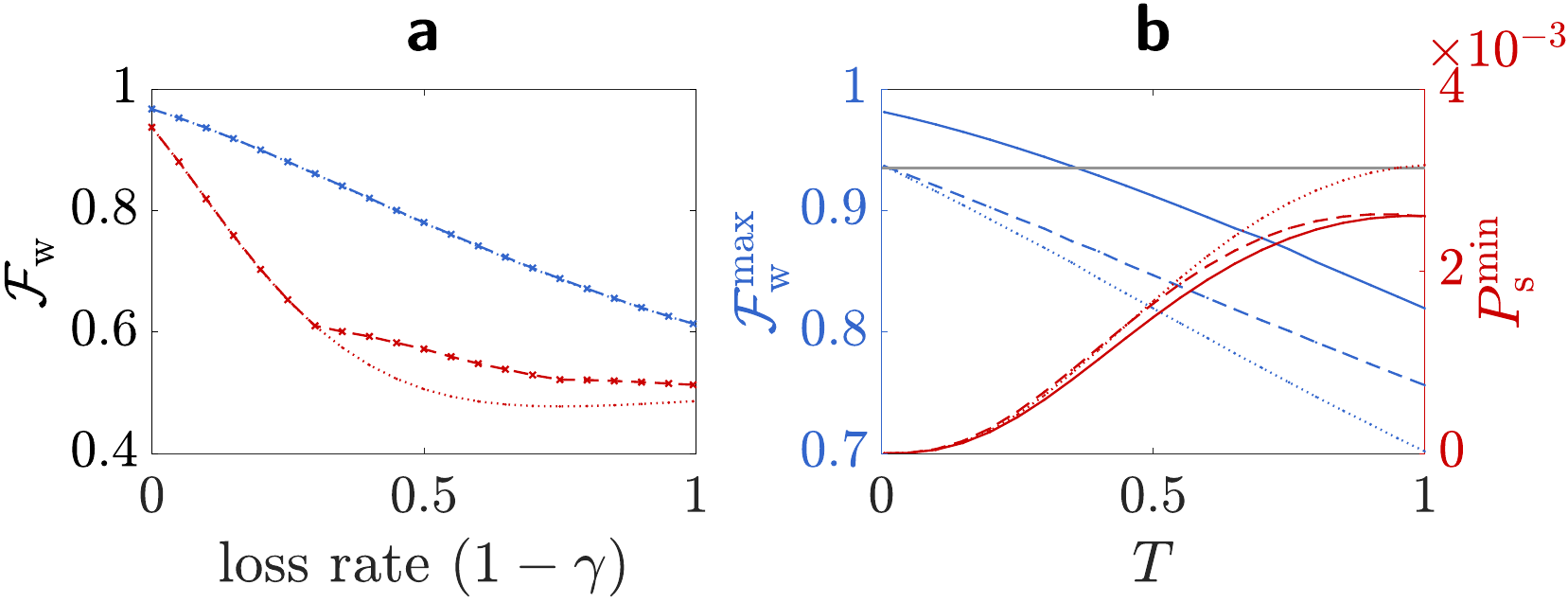}
\caption{(a) Worst-case fidelity (over all pure encoded states) for $(\overline{0},\overline{1})$ encoding with the recovery map from Ref. \cite{bergmann2016quantum} (dotted) and $(\overline{+},\overline{\vphantom{+}-})$ encoding with numerically optimized recovery maps (dashed) versus the loss rate for $T=35\%$. The choice of $T$ is motivated by (b) optimized worst-case fidelity (blue) and the minimum success probability (red) plotted against $T$ for the loss rate of $10\%$ with $\eta=90\%$ (dotted), the loss rate of $5\%$ with $\eta=90\%$ (dashed), and the loss rate of $10\%$ with $\eta=95\%$ (solid) and where the solid gray line represents the worst-case fidelity benchmark of $93.5\%$ resulting from Ref.~\cite{bergmann2016quantum} wherein ideal amplification using teleportation after recovery is considered instead of noisy teleamplification before recovery.}
\label{fig: main-results}
\end{figure}
\begin{figure}[hp]
\centering
\includegraphics[width=10cm]{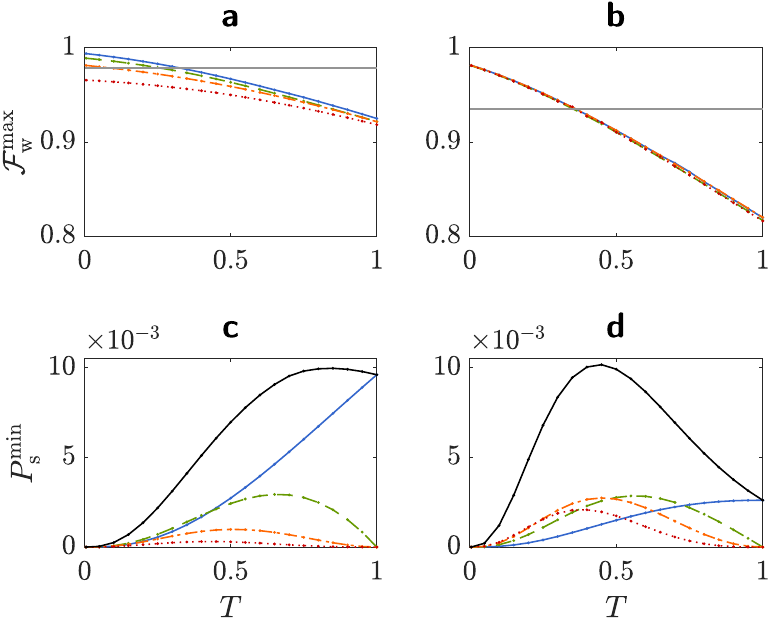}
\caption{Optimized worst-case fidelities for (a) $\alpha=\sqrt{2}$ and (b) $\alpha=2$. The worst-case fidelity due to Ref. \cite{bergmann2016quantum} is shown by the horizontal gray line. Minimum success probabilities for (c) $\alpha=\sqrt{2}$ and (d) $\alpha=2$. The overall minimum success probability of accepting up to 10 photons in each of the relevant output modes and adapting the recoveries appropriately is depicted in black. These are plotted against the beamsplitter transmittance $T$ for a loss rate of $10\%$ and a detector efficiency of $\eta=95\%$. The results for effective ZPS, 1PS, 2PS, and 3PS are shown in blue, green, orange, and \mbox{red, respectively.} For $\alpha=\sqrt{2}$, ZPS outperforms the others in terms of the fidelity. On the other hand, sufficiently large initial and final amplitudes such as $\alpha\gtrsim 2$, the values for any multiphoton subtraction lie close to \mbox{each other.}}
\label{fig: compare-nPS}
\end{figure}
\begin{figure}[p]
\centering
\includegraphics[width=7cm]{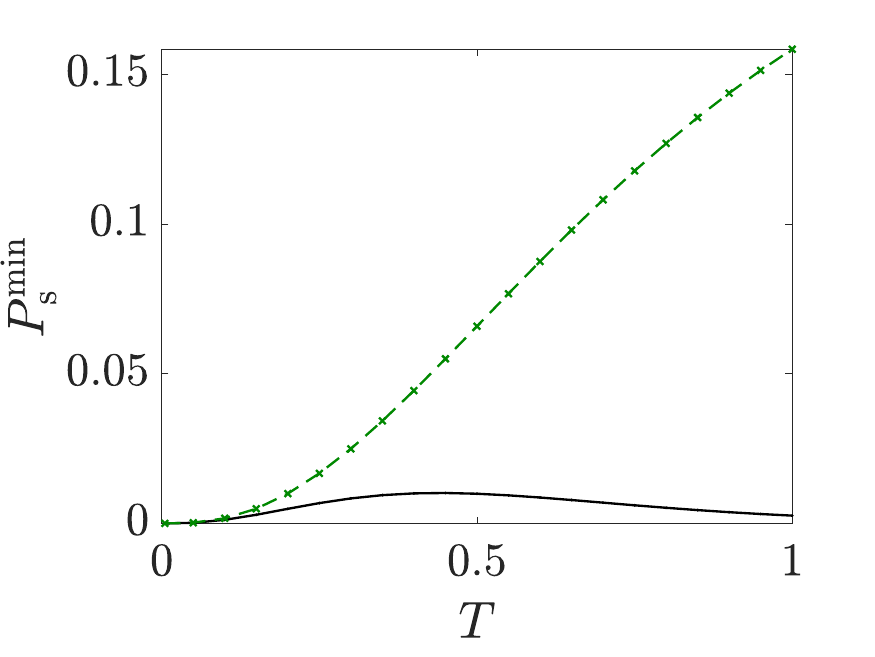}
\vspace{-10pt}
\caption{The minimum success probability considering subtractions of up to 10 photons for noiseless attenuation and up to 10 photons in each of the relevant ancilla modes for teleamplification is plotted in green (dashed). The solid black curve is the same as in Fig. \ref{fig: compare-nPS}(d) and corresponds to accepting up to 10 photons for noiseless attenuation but with teleamplification heralded only on the coincident measurement of single photons in the relevant modes. All parameters are the same as in Fig. \ref{fig: compare-nPS}(b,d). The large improvement in the probability of success is because we accept a larger number of compatible outcomes to herald the outcome and adapt the recoveries accordingly.}
\label{fig: increase-PS-teleamp}
\end{figure}

Attenuating the coherent-state amplitudes in encoded qubits also reduces uncorrectable errors caused by detector inefficiencies during teleamplification. For smaller coherent-state amplitudes, the fidelity without suppression has higher values, as expected. However, a smaller amplitude does not benefit quantum information tasks in later applications. Noiseless attenuation is directly controlled by the beamsplitter transmittance, and lower values lead to a dramatic increase in fidelity even if start with a large amplitude. Any nonzero amplitude can be amplified during the noiseless amplification stage of the scheme to produce a well-defined outcome. 

The recovery operations used in Fig. \ref{fig: full-setup-effectiveness} are suboptimal when channel losses are large. This is seen by the large difference among the plots (dashed-red and dotted-red) for when noiseless attenuation is absent, \mbox{in Fig. \ref{fig: main-results}(b).} However, noiseless attenuation before loss still shows a significant improvement for realistic system parameters for such large channel losses, even after considering \mbox{optimal recoveries.} Note that for optimizing the recoveries, we transition to using the $(\overline{+},\overline{\vphantom{+}-})$ basis for its operational convenience, as explained in Sec. \ref{sec: optimal-recovery}.

As we saw earlier, the greatest improvement in fidelity occurs when the beamsplitter transmittance is chosen near zero. However, Fig. \ref{fig: main-results}(b) shows that it comes at the expense of the probability of success. A benchmark in fidelity of about $93.5\%$ is set by the setup without any noisy teleamplification while also considering perfect amplification after recovery as in Ref. \cite{bergmann2016quantum}. 

To go beyond this benchmark, the minimum success probability (over all input pure states) takes the value of $0.08\%$ for cat codes with an initial coherent-state amplitude $\alpha=2$, detector efficiency of $95\%$, and the environmental-loss rate of $10\%$. This occurs for beamsplitter transmittance of about $T=35\%$ used for the photon subtraction.

The worst-case fidelity naturally does not overcome such a threshold even if it is combined with a smaller overall channel loss for sufficiently small \mbox{detector efficiency.} This is explained by the high additional noise from inefficiencies while trying to address the noise from the channel~loss.

 The success probability of the procedure is improved by accepting the subtractions of arbitrary $n$ photons from the encoded qubits as shown by the Figs. \ref{fig: compare-nPS}(c, d). In the $(\overline{+},\overline{\vphantom{+}-})$ basis, the distortion of the qubit due to redistribution of the probability amplitudes is undone when we use ZPS with standard teleamplification. However, when we accept any arbitrary number of photon subtraction, it can cause a nonunitary distortion even after restoring the coherent-state amplitudes. This becomes prominent when the initial coherent-state amplitude is small and the differences in the worst-case fidelities are seen clearly in Fig. \ref{fig: compare-nPS}(a) for $\alpha=\sqrt{2}$. The optimized CPTP recovery channels cannot restore a distorted qubit with high fidelity when $n\,(\textrm{mod}\,4)\neq0$. 
 
 On the other hand, for sufficiently large initial (and therefore also the final) coherent-state amplitudes $\alpha\gtrsim2$, the fidelities under different amounts of photon subtraction are roughly the same, as the distortions are neglible, as seen by comparing the curves in Fig. \ref{fig: compare-nPS}(b). Therefore, all arbitrary photon subtraction can be accepted to increase the success probability to about $0.94\%$.

The results discussed have been restricted to standard teleamplification so far. However, all combinations of $ x_k\neq0$ can be accepted for large cat states with large coherent-state amplitudes and recovery is appropriately adapted to further increase the success rate. These combinations also herald a vacuum in mode $A$, and due to the symmetry of the encoding, we obtain one of the same four plots for the worst-case fidelities for all four cases corresponding to $n' = (n+x_1+x_2+x_3-3)(\textrm{mod}\,4)$  \mbox{shown by Fig. \ref{fig: compare-nPS}.} 

\pagebreak
 Accepting up to 10 photons in each relevant mode such that $x_k\neq0$ increases the previously discussed success probability to about $3.42\%$, as illustrated by Fig. \ref{fig: increase-PS-teleamp}. With this sufficiently large success rate the procedure becomes a promising alternative for error suppression for passive photon losses. 
 
 Note that all the outcomes with up to 7 photons detected, including in the mode $A$, constitute more than $99.6\%$ of all possible outcomes for the $\smlket{\overline{+}}$ state as an example, making it a good approximation to cutoff at \mbox{10 photons.}
\section{Discussion}
\label{sec: discussion}

  In this paper, we analyze noiseless loss suppression in the context of error-correcting codes. In particular, we examine the four-component cat codes and how well they can be protected against uncorrectable errors resulting from decoherence from an amplitude-damping channel  prior to error correction. This is typically known as error suppression.

 Sufficiently large amplitude cat states such as those with the coherent-state amplitude of 2, can be error corrected using noiseless recovery maps to a worst-case fidelity of about $93.5\%$ under a significantly large amplitude-damping channel of $10\%$. We show worst-case fidelities beyond that which can be achieved solely by error correction, using realistic and therefore noisy detectors at the cost of success probability of about $3\%$. 

  Qubits encoded on the cat codes require $\alpha\gtrsim2$ coherent-state amplitude for efficient quantum computation tasks and the amplitude-damping channel reduces this amplitude making noise-free amplification essential to amplitude restoration. 
  However, large-amplitude coherent states at the input are disturbed more by the lossy channel as they carry more energy, and so it is reasonable that noiselessly attenuating the signal before any transmission through the lossy channel reduces the effects of decoherence and therefore, also suppresses uncorrectable logical errors on the encoded states. 
  
  However, noiseless attenuation and amplification required for such loss suppression are implemented nondeterministically with photon subtraction and teleamplification, respectively, for the cat states. These processes require conditional measurements with photon-number-resolving detectors which introduces further noise into the system. Therefore, the impact of these on the effectiveness of the overall scheme needs to be assessed with realistic values of detector~efficiency. 
  We found that with currently available superconducting-nanowire detectors, noiselesss loss suppression can lead to substantial gains in optimized worst-case fidelity over all the input pure~states. 
  
  The ability to protect large cat states under large losses is quite useful for quantum information applications in the near term when such losses before error correction are unavoidable in a physical realization even if we assume noise-free encoding, error correction, and decoding~procedures. Near-term applications include storing photonic qubits in a cavity or long fiber loops, and for direct communication of encoded qubits over long distances as a stepping-stone towards distributed quantum~computing~\cite{azuma2023quantum}.
    
  For the future outlook, it may be interesting to extend the analysis presented here to cat codes with more than four components, such as in Refs. \cite{bergmann2016quantum, guanzon2023noiseless}, arbitrary rotation symmetric codes \cite{grimsmo2020quantum, endo2022quantum} and translation symmetric codes such as GKP codes \cite{gottesman2001encoding}. Similar preemptive measures could be sought for other relevant noise models, like the dephasing channel, such as in Refs. \cite{gisin2005error, lee2023error} that can affect the photonic qubits. 
  
  It may be interesting to study loss suppression on arbitrary nongaussian, nonclassical states with other nongaussian conditional measurements photon catalysis, for example, that have been useful in quantum key distribution \cite{lvovsky2002quantum, hu2017continuous, ma2018continuous, singh2021non, jafari2024long}. More generally, better error suppression techniques can be explored for arbitrary noise channels on diverse platforms \cite{endo2022quantum}. Finally, the nonconvex optimization algorithms used to find the maximum worst-case fidelity may be improved. 

\newpage
\begin{backmatter}
\bmsection{Acknowledgments}
S.U. Shringarpure acknowledges insightful conversations with Jaehak Lee and Soumyakanti Bose. 
The authors acknowledge support from the National Research Foundation of Korea (NRF) grants funded by the Korean government (Grant Nos. NRF2020R1A2C1008609, 2023R1A2C1006115, NRF2022M3E4A1076099 and RS5 2023-00237959) via the Institute of Applied Physics at Seoul National University, the Institute of Information \& Communications Technology Planning \& Evaluation (IITP) grant funded by the Korea government (MSIT) (IITP-2021-0-01059 and IITP-20232020-0-01606), and the Brain Korea21 FOUR Project grant funded by the Korean Ministry of Education.

\bmsection{Disclosures}
The authors declare no conflicts of interest.

\bmsection{Data availability} 
No data were generated or analyzed in the presented research. The results in this paper can be reproduced using the files in \url{github.com/saurabh-shringarpure/nps-teleamp}.

\end{backmatter}

\appendix
\section{Equivalent setups for noisy multiphoton subtraction, teleamplification and their combination}
\label{append_sec: equivalent-setups}

\subsection{Multiphoton subtraction}
Noisy $n$-photon subtraction with a beamsplitter of transmittance $T$ generates,
\begin{align}
    \begin{split}
        &e^{-\frac{\abs{\alpha}^2\eta R}{2}}
        \frac{\argp{\sqrt{\eta R} \alpha}^n}{\sqrt{n!}}
        \sum_{k=0}^{3}{c_k u^{nk}
        \ket{\sqrt{T}\alpha_k}
        \ket{\sqrt{(1-\eta)R}\alpha_k}_{\rm{env.}}}. 
    \end{split}
\end{align}

Alternatively, consider a setup where the encoded state first experiences  noiseless amplification by $\mu^{\hat{n}}$ followed by an ideal $n$-photon subtraction with the same beamsplitter and lastly a loss of $(1-\eta')$ to give
\begin{equation}
        e^{-\frac{\abs{\alpha}^2(1-\mu^2 T)}{2}}
        \frac{\argp{\sqrt{R\mu^2}\alpha}^n}{\sqrt{n!}}
        \sum_{k=0}^{3}c_k u^{nk} 
        \ket{\sqrt{\eta'T\mu^2}\alpha_k}
        \ket{\sqrt{(1-\eta')T\mu^2}\alpha_k}_{\rm{env.}}.
\end{equation}
Inspection shows that the two output states become equivalent, up to a factor, when
$\mu = {1}/{\sqrt{\eta'}}$ \mbox{and $\eta' = {T}/\argp{1-\eta R}$.} The attenuation is the same as for ZPS. The difference is only in the prefactor and the phases $u^{nk}$ attached to the various coherent-state amplitudes. 

If the output state of a noisy ZPS passes through an amplitude-damping channel with transmittance $\gamma$ the effective transmission becomes the product $\eta'\gamma$. Therefore, we can analyze the performance of cat codes modified by noisy ZPS by using Eq. (\ref{eq: rho}) and these modified values of the initial amplitude and channel transmittance.

\subsection{Teleamplification}
\begin{figure}[t]
\centering
\includegraphics[width=12cm]{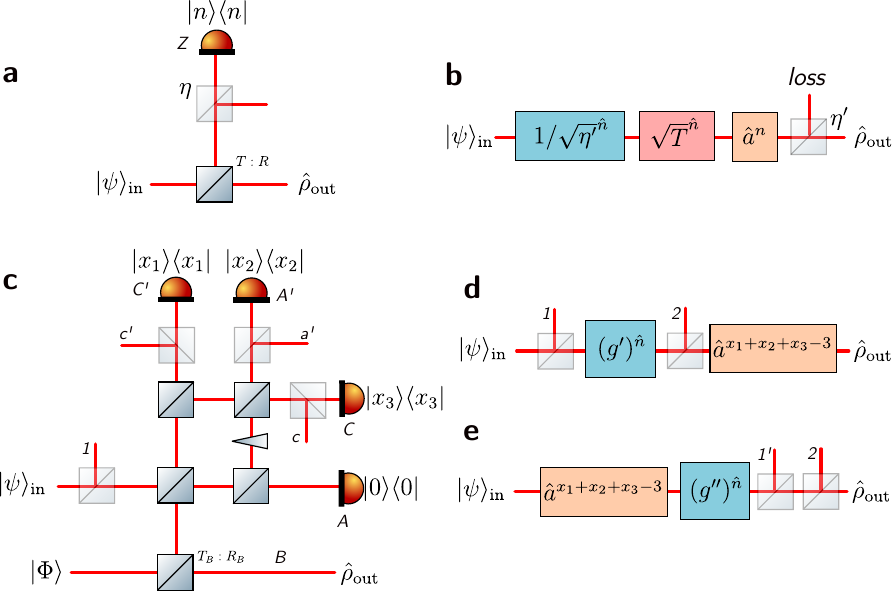}
\caption{An inefficient detector is modeled by placing a beamsplitter with transmittance $\eta$ before a \mbox{perfect detector.} (a) Noisy $n$-photon subtraction with a photon-number-resolving detector of efficiency $\eta$ is equivalent, up to a factor, to (b) a noiseless preamplification by $1/\sqrt{\eta'}$, noiseless attenuation and $n$-photon subtraction followed by a loss \mbox{of $1-\eta'$.} Similarly, (c) noisy teleamplification is equivalent, up to a constant factor, to (d) a  noiseless amplification by $g'=g/\sqrt{1-R_2}$ followed by some loss $R_2$ and the subtraction of \mbox{$(x_1+x_2+x_3-3)$ photons.} The input loss denoted by $R_1$ is unchanged. (e) The setup with input losses before amplification can be made again equivalent to photon subtraction followed by amplification by $g''$, and the losses $R_1'$ and $R_2$. Such manipulations are possible due to cat encoding of the input states.}
\label{fig: equivalent-setups}
\end{figure}

The result of non-ideal teleamplification assuming that we have detectors with efficiency $\eta$ that measure, say, $(x_1,x_2,x_3)$ number of photons in modes $C$, $A'$ and $C'$, and a loss $R_1$ after the preparation of the input state, as shown in Fig. \ref{fig: equivalent-setups}(c), is straightforward with an unchanged $g$ and $\beta = \sqrt{{(1-R_1) }/{R_B}}\alpha.$

Equation (\ref{eq: ideal teleamp}) when $j=k$ is extended as
\begin{align}
\begin{split}
    &\mathcal{M}_{+}^{'}\ket{0}_A\sum_{k=0}^{3} c_k u^{-3k}\ket{g\sqrt{1-R_1}\alpha_k}_B 
    \ket{-\sqrt{R_1}\alpha_k}_1
    \ket{-\sqrt{\eta(1-R_1)} \alpha_k}_C \ket{\sqrt{(1-\eta)(1-R_1)} \alpha_k}_{c}
    \\
    &\ket{\sqrt{\frac{\eta}{2}(1-R_1)} \alpha_k}_{A'} \ket{-\sqrt{\frac{1-\eta}{2}(1-R_1)} \alpha_k}_{a'}
    \ket{\sqrt{\frac{\eta}{2}(1-R_1)} \alpha_k}_{C'} \ket{-\sqrt{\frac{1-\eta}{2}(1-R_1)} \alpha_k}_{c'},
\end{split}
\end{align}
where the additional modes $1$, $c$, $a'$ and $c'$ are lost to the environment and ideal projections onto $(x_1,x_2,x_3)$ Fock states in the modes $C$, $A'$ and $C'$ can be assumed.

We can also combine all losses into a single mode, $2$, as shown in Fig. \ref{fig: equivalent-setups}(d), and measure the modes $A$, $C$, $A'$, and $C'$, such that the unnormalized state is given by

\begin{align}
\begin{split}
    &\mathcal{M}_{+}^{'} 
    e^{-\eta(1-R_1)\alpha^2}\frac{\argp{\sqrt{\eta(1-R_1)}\alpha}^{x_1+x_2+x_3}}{\sqrt{x_1!x_2!x_3!2^{x_2+x_3}}}\times
    \\
    &\sum_{k=0}^{3}c_k u^{k(x_1+x_2+x_3-3)}\ket{-\sqrt{R_1}\alpha_k }_1 
   \ket{\sqrt{1-R_2}g'\sqrt{1-R_1}\alpha_k}_B
    \ket{-\sqrt{R_2}g'\sqrt{1-R_1}\alpha_k }_2,
\end{split}
\end{align}

where
\begin{align}
    g'= \frac{g}{\sqrt{1-R_2}}\ {\rm {and}}\ 
    R_2=\frac{2(1-\eta)}{g^2+2(1-\eta)}.
 \end{align}
\newpage
\subsection{The combined setup}

 The output of the ideal attenuation, from Fig. \ref{fig: equivalent-setups}(b), can be taken as input to the teleamplification in Fig. \ref{fig: equivalent-setups}(c), and the effective losses during and after $n$-photon subtraction can be represented by input losses with substitution $R_1 \mapsto 1-\eta'\gamma.$

Furthermore, consider a situation where these input losses and the  noiseless amplification are switched in the order shown in Fig. \ref{fig: equivalent-setups}(e) for which we need 
solutions to
\begin{align}
\begin{split}
\sqrt{R_1}=g''\sqrt{R_1'}\ {\rm and}\ 
g'\sqrt{1-R_1}=g''\sqrt{1-R_1'},
\end{split}
\end{align}
which are
\begin{align}
\begin{split}
g''= \sqrt{(1-R_1)g'^2+R_1}\ {\rm and}\ 
R_1'= \frac{R_1}{(1-R_1)g'^2+R_1}.    
\end{split}
\end{align}

We have a constraint,
\begin{equation}
    \sqrt{{T}/{\eta'}} (g''){\sqrt{(1-R_1')(1-R_2)}} = 1,
\end{equation}
to restore the coherent-state amplitude of the encoded qubit which gives the equation, 
\begin{align}
\begin{split}
g^6&(\gamma T)^2 +
g^4\gamma  T \argc{\gamma  T-\eta \args{(2 \gamma -1) T+1}}+
\\
g^2&\argc{\eta -T \args{\eta +\gamma  (1-2 \eta) }-1} 
+2\args{1-\eta} \args{\gamma T  +\eta(1- T) -1} =0.        
\end{split}
\end{align}
It is easily solved for real-valued $g\geq1$ in \mbox{\textsc{mathematica}}. The final state is then simply given by Eq. (\ref{eq: rho}) with substitutions
\begin{equation}
T \mapsto T_{\rm {eff.}}={Tg''^2}/{\eta'} ,
\,\gamma \mapsto \gamma_{\rm {eff.}}=(1-R_1')(1-R_2),
\label{eq: effective T and gamma}
\end{equation}
and the attenuation with $n$-photon subtraction and and a general teleamplification with heralding on $(x_1,x_2,x_3)$ in the modes $C$, $A'$, and $C'$ can be put together in a complete setup. 

The success probability, for an input $\sum_{k=0}^{3} c_k \ket{\alpha_k}$ \mbox{becomes}
\begin{align}
    \begin{split}
        P_{\rm{s}}=&\args{
        \mathcal{M}_{+}^{'}
        e^{-\eta\argp{\frac{R}{2}+\gamma T}\alpha^2}
        \frac{\argp{\sqrt{\eta R}\alpha}^n\argp{\sqrt{\eta \gamma T}\alpha}^{x_1+x_2+x_3}}{\sqrt{n!x_1!x_2!x_3!}\sqrt{2}^{x_2+x_3}} 
        }^2\times
        \\
        &\sum_{j,k=0}^{3}c_j^*u^{(-j+k)(n+x_1+x_2+x_3-3)}c_k 
        e^{-\alpha^2\args{1-\eta R+\gamma T\argp{1+g^2-2\eta}}\argp{ 1- u^{-j+k} }},
    \end{split}
    \label{eq: Ps-zps-teleamp}
\end{align}
where $\mathcal{M}_{+}^{'}$ is a function of the coherent-state \mbox{amplitude $\beta = \sqrt{(1+g^2)\gamma T}\alpha$.}

\section{Alternate setup for amplitude restoration}
\label{append_sec: alternative}
In the main text, we considered the use of teleamplification to restore the coherent-state amplitude after the loss channel. Alternatively, the loss channel may be placed within the teleamplification setup, in the mode that contains one part of the entangled resource state. This is the section of mode $C$ in Fig. \ref{fig: setup} between modes $A$ and $B$. 

\begin{figure}[htp]
\centering
\includegraphics[width=6cm]{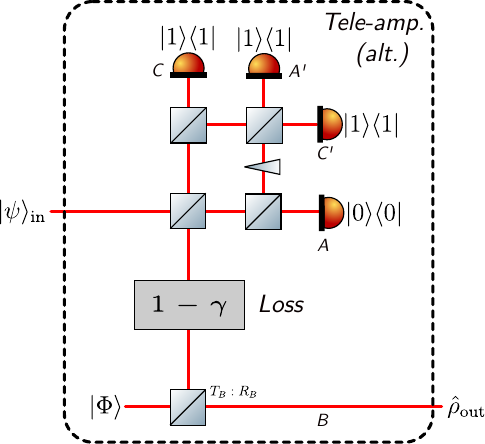}
\caption{An alternate approach to restore the inevitable drop in coherent-state amplitude due to the loss channel. Half of the entangled resource state is sent through the loss channel to interact with the encoded qubit.}
\label{fig: setup_alt}
\end{figure}
   
Assume perfect detectors and no ZPS for simplicity to compare the performance of this alternate setup to the original one in Fig. \ref{fig: setup}. Losses occur only through the loss channel with transmittance $\gamma$.
Following the procedure described before, we get the heralded state
\begin{align}
    \begin{split}
        \sim &\sum_{k=0}^{3} c_k \ket{g_{\rm {alt.}} \alpha_k}_B \ket{\sqrt{\frac{1-\gamma}{\gamma}}\alpha_k}_{\rm{env.}},
    \end{split}
\end{align}
where $g_{\rm{alt.}} = 1 = \sqrt{{T_B}/{R_B\gamma}}$
as opposed to the heralded state in the original setup
\begin{align}
    \begin{split}
        \sim \sum_{k=0}^{3} c_k \ket{g_{\rm{ori.}} \sqrt{\gamma} \alpha_k}_B\ket{-\sqrt{1-\gamma}\alpha_k}_{\rm {env.}}
    \end{split}
\end{align}
where $g_{\rm{ori.}} = {1}/{\sqrt{\gamma}} = \sqrt{{T_B}/{R_B}}.$
 
The alternate setup demands higher values of the gain from teleamplification to restore the coherent-state amplitude as the losses grow. This is due to the factor of $\gamma$ in the denominator. 

It in turn leads to the demand for a larger coherent-state amplitude in the resource state. This amplitude has values $\beta_{\rm {alt.}} = \sqrt{{1}/{R_B\gamma}}\alpha$ and $\beta_{\rm{ori.}} = \sqrt{{\gamma}/{R_B}}\alpha,
$
in the two setups.

Compare the intensities lost to the environment in the two cases corresponding to any $\alpha_k$. For the original setup, it is $(1-\gamma)\alpha^2$, which grows linearly with the losses. On the other hand, it is $(1-\gamma)\alpha^2/\gamma$ for the alternate one and blows up as the losses increase.

Therefore, the original setup used in the main text is more resilient to the loss channel. The loss of entanglement in the resource along with the requirement of a larger coherent-state amplitude for the resource state negatively impacts this setup more, reducing coherence.

\section{Numerical optimization}
\label{append_sec: numerical-opt}

\subsection{CP, TP, and CPTP projections in terms of the Choi matrix for initiating the optimization}
\label{append_sec: CPTP init}

Initiating the optimization requires a random, completely-positive, trace-preserving (CPTP) matrix. This is constructed from a random full-rank \textit{seed} matrix that is subjected to a projection onto the space of CPTP matrices \cite{boyd_vandenberghe_2004, plesnik2007finding, knee2018quantum, ahmed2023gradient}.

The CPTP projection involves alternating CP and TP projections according to Ref. \cite{knee2018quantum}. The CP projection, individually, is obtained by simply replacing the negative eigenvalues of a matrix to zero in its eigendecomposition, i.e.
\begin{equation}
    \mathcal{CP}\args{\mathcal{R}} = V \max\argp{D, 0} V^{-1}, 
\end{equation}
where $\mathcal{R} = VD V^{-1}$ is the eigendecomposition.

The TP projection is given by
\begin{equation}
    \mathcal{TP}\args{\mathcal{R}}= {\rm {vec}}^{-1}\left[\sket{\mathcal{R}}-\frac{1}{{\rm{dim}}\mathcal{K}}M^\dagger M \sket{\mathcal{R}}+\frac{1}{{\rm{dim}}\mathcal{K}}M^\dagger \sket{I_{\mathcal{H}}} \right],
\end{equation}
and 
\begin{equation}
    M = \sum_k {I_{\mathcal{H}} \otimes \bra{k} \otimes I_{\mathcal{H}} \otimes \bra{k}},
\end{equation}
where $\ket{k}$ are the appropriate orthonomal bases in $\mathcal{K}$.

\subsection{The gradient of fidelity}
\label{append_sec: grad-fidelity}

We are interested in the fidelity of the recovered qubit after the channel $\mathcal{E}$ (Eq. (\ref{eq: effective-channel})) to the input one, a pure state in the computational basis given by
\begin{equation}
    \mathcal{F} = {\rm {tr}}\argc{ \smlketbra{\psi}{\psi} \mathcal{E}\args{\smlketbra{\psi}{\psi}}  }.
\end{equation}
Assuming ideal encoding, recovery, and decoding operations, it may be rewritten in terms of the Choi matrix for the recovery operation as
\begin{equation}
    \mathcal{F} = {\rm {tr}}_{\mathcal{K}}\argc{ \rho\;{\rm {tr}}_{\mathcal{H}}\argc{\argp{\rho'^\intercal\otimes I_{\mathcal{K}}}\mathcal{R}}}= {\rm {tr}}\argc{\argp{\rho'^\intercal\otimes  \rho} \mathcal{R}},
\label{eq: fidelity}
\end{equation}
where $\rho$ is input, pure state after encoding and $\rho'$ is the state just before the recovery.

Therefore, the variation of fidelity becomes
\begin{align}
    \delta \mathcal{F} = {\rm {tr}}\argc{\argp{\rho'^\intercal\otimes  \rho} \delta \mathcal{R}}
    ,
\label{eq: variation-of-fidelity}
\end{align}
and the gradient in $\mathcal{R}$,
\begin{equation}
    \mathcal{M} = \rho'^\intercal\otimes  \rho.
\end{equation}

\subsection{CPTP constrained variation of Choi matrices}
\label{append_Sec: CPTP-deltaR}

Consider a CPTP channel represented by Choi matrix $\mathcal{R}$. A small unrestricted variation on this channel under the constraint that it remains CPTP, is given by \cite{teo2011adaptive}

\begin{equation}
    \mathcal{R}+\delta\mathcal{R} = \argp{I+\delta Z^\dagger}\mathcal{R}\argp{I+\delta Z},
    \label{eq: deltaR}
\end{equation}
where $\delta Z$ is an operator such that the TP constraint ${\rm{tr}}_{\mathcal{K}}\argc{\mathcal{R}} = I_{\mathcal{H}}$
is satisfied where $\mathcal{K}$ and $\mathcal{H}$ are the output and the input  Hilbert spaces of the channel. The matrix $\delta Z$ is further parametrized by a small, arbitrary Hermitian operator, $\delta{A}$, such that the CP constraint is upheld, which gives \cite{teo2011adaptive}
\begin{equation}
    I+\delta Z = (I+\delta{A})\args{\sqrt{ {\rm{tr}}_\mathcal{K} \argc{\argp{I+\delta{A}^\dagger}\mathcal{R} \argp{I+\delta{A}}}} \otimes I_{\mathcal{K}}}^{-1}.
    \label{eq: CPTP parameterization}
\end{equation}

\subsection{CPTP restricted steepest ascent}
\label{append_sec: deltaA}

Consider a function $\mathcal{F}$ of a CPTP-constrained argument $\mathcal{R}$ whose variation is given by 
\begin{equation}
    \delta{\mathcal{F} } = {\rm {tr}} \argc{\mathcal{M} \delta{\mathcal{R}}},
\end{equation}
where $\mathcal{M} $ is its gradient in $\mathcal{R}$.

A small CPTP restricted ascent in $\mathcal{F}$ in the \textit{direction} of the highest gradient can be constructed using Eqs. (\ref{eq: deltaR}) and (\ref{eq: CPTP parameterization}) with \cite{teo2011adaptive}
\begin{equation}
    \delta{A} = \delta{A}^\dagger = \frac{\epsilon}{2}\argp{\mathcal{M} -\frac{1}{2}{\rm {tr}}_{\mathcal{K}}\argc{\mathcal{M} \mathcal{R} +\mathcal{R} \mathcal{M}}\otimes I_{\mathcal{K}}},
\end{equation}
for some small $\epsilon>0$.

\subsection{A typical instance of numerical optimization}

\begin{figure}[htbp]
\centering
\includegraphics[width=10cm]{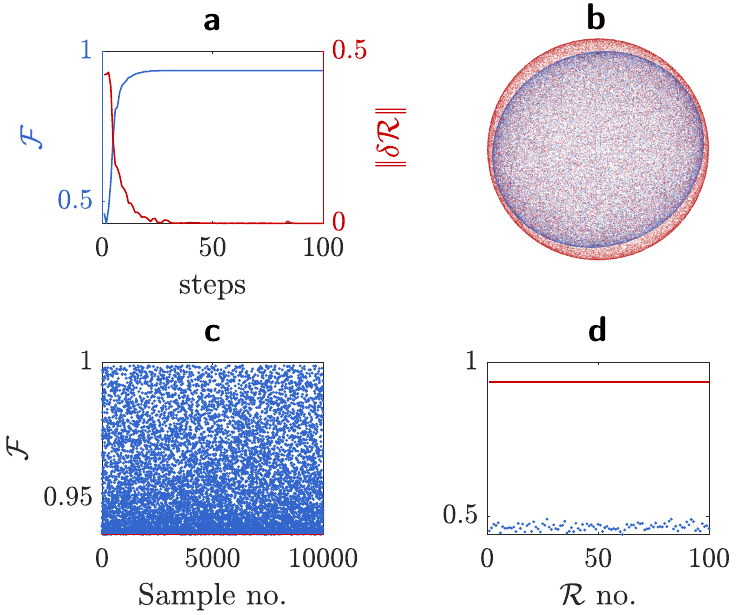}
\caption{An instance of optimization for the optimal fidelity in Fig. \ref{fig: main-results}(a) with ZPS for $T=35\%$ and $10\%$ loss rate with $\eta=95\%$.
(a) The operator norm of the variation on the recovery map (red), $\delta\mathcal{R}$, and the optimized fidelity (blue) are plotted against the step count. Fidelities saturated within 100 steps for the initial $\epsilon=3$, which is decreased by $0.994$ at each step. (b) Recovered qubit states (blue) resulting from the optimized recovery map (after 1000 steps), derived from 50,000 uniformly sampled initial pure states on a Bloch sphere (red). (c) Fidelities of the optimally recovered qubits to 10000 uniformly sampled input states are shown in blue. These are above the red line that represents the worst-case fidelity $93.5\%$ for the optimal $\mathcal{R}$. This indicates that the optimized solution is indeed the worst-case fidelity for a given recovery. (d) Worst-case fidelity of $100$, full-rank, unoptimized, randomly chosen channels $\mathcal{R}$ in blue. These lie below the red line, supporting that the optimized recovery map performs better.}
\label{fig: opt-instance}
\end{figure}

The results for an instance of numerical optimization for the results shown in Fig. \ref{fig: main-results}(a) are presented \mbox{in Fig. \ref{fig: opt-instance}.} The final optimized recovery map is block-diagonal, with blocks corresponding to the subspaces with even and odd photon-number parities.

\end{document}